\documentclass[a4paper,11pt]{article} \usepackage{pos} \usepackage{enumitem}
\usepackage{slashed} \usepackage{acronym} \usepackage{menukeys}
\usepackage[capitalize]{cleveref} \newcommand{\editframe}{\textit{edit frame}}
\newcommand{\mainframe}{\textit{main frame}}
\newcommand{\modelfile}{\textit{model file}}

\newcommand{\myacrodef}[3]{\newcommand{#1}{\abbrev{#2}}}

\newcounter{notecount}

\myacrodef{\QED}{QED}{Quantum Electrodynamics}
\myacrodef{\qcd}{QCD}{Quantum Chromo Dynamics}
\myacrodef{\lhc}{LHC}{Large Hadron Collider}
\myacrodef{\lo}{LO}{leading order}
\myacrodef{\nlo}{NLO}{next-to-leading order}
\myacrodef{\nnlo}{NNLO}{next-to-next-to-leading order}
\myacrodef{\llog}{LL}{leading logarithmic}
\myacrodef{\pdf}{PDF}{parton density function}
\myacrodef{\sm}{SM}{Standard Model}
\myacrodef{\bsm}{BSM}{beyond-the-\ac{SM}}
\myacrodef{\mssm}{MSSM}{Minimal Supersymmetric \ac{SM}}
\myacrodef{\susy}{SUSY}{Supersymmetry}
\myacrodef{\dreg}{DREG}{Dimensional Regularization}
\myacrodef{\dred}{DRED}{Dimensional Reduction}
\myacrodef{\emt}{EMT}{energy-momentum tensor}
\newcommand{\keystroke}[1]{\keys{#1}}

\newcommand{\abbrev}[1]{{\scalefont{.9}#1}}

\newcommand{\dd}{\mathrm{d}}
\newcommand{\deriv}[3]{\frac{\partial\ifthenelse{\equal{#1}{}}{}{^{#1}}%
    #2}{\partial #3\ifthenelse{\equal{#1}{}}{}{^{#1}}}}
\newcommand{\dderiv}[3]{\frac{\dd\ifthenelse{\equal{#1}{}}{}{^{#1}}%
    #2}{\dd #3\ifthenelse{\equal{#1}{}}{}{^{#1}}}}

\newcommand{\feyngame}{\texttt{FeynGame}}

\title{FeynGame-2.1 -- Feynman diagrams made easy}
\author*{Robert Harlander}
\author{Sven Yannick Klein}
\author{Magnus Schaaf}

\affiliation{TTK, RWTH Aachen University,
  Sommerfeldstr.~16, 52056 Aachen, Germany}

\emailAdd{robert.harlander@rwth-aachen.de}
\emailAdd{sven.yannick.klein@rwth-aachen.de}
\emailAdd{magnus.schaaf@rwth-aachen.de}

\abstract{\feyngame\ is an open-source software tool to draw Feynman diagrams,
  but also to get acquainted with their structure. This article reports on a
  number of new features which have been added to \feyngame\ since its first
  release. These include full support of \LaTeX\ for the line and vertex
  labels, the possibility to automatically include momentum arrows, new
  graphical elements, and new pedagogical features. \feyngame\ is freely
  available
  \begin{itemize}
  \item 
    as \texttt{jar} or MacOS \texttt{app} file from
    \url{https://web.physik.rwth-aachen.de/user/harlander/software/feyngame}
  \item as source code from \url{https://gitlab.com/feyngame/FeynGame}
  \end{itemize}
}

\FullConference{The European Physical Society Conference on High Energy Physics (EPS-HEP2023)\\
 21-25 August 2023\\
Hamburg, Germany\\}

\begin{document}
\maketitle

%- }}}
%- {{{ section{Introduction}

\section{Introduction}\label{header}

\feyngame\ is a Java tool for drawing and playing with Feynman diagrams. The
initial idea for its development was to provide a tool for high-school
students, teachers, or undergraduate students which allowed them to get
familiar with the concept of Feynman diagrams in a playful way. While this is
still one of the main purposes of \feyngame, its functionality allows one to
use it as a simple, efficient, and flexible drawing tool for Feynman diagrams.

The main feature which distinguishes \feyngame\ from other drawing tools is
that it is based on particle physics models. This means that
\feyngame\ ``knows'' about the Feynman rules of a particular theory. This
information is provided to \feyngame\ in the form of a \modelfile.
By default, \feyngame\ assumes the Standard Model (\sm) as the underlying theory, and the user
may start from the corresponding \modelfile\ to build \modelfile s for simpler
or more elaborate theories.

This model-based approach has two important consequences. On the one hand,
every particle of a particular theory may be given unique attributes, such as
the line style or a text label. For example, a gluon may be represented by a
red spiral line and the label $g$, a Higgs boson by a dashed blue line and the
label $H$. This makes the drawing of Feynman diagrams very efficient.  Assume
that one would like to distinguish top-quark lines in a Feynman diagram from
other fermion lines by drawing them a bit thicker and/or in a different
color. As opposed to other Feynman diagram drawing tools, there is no need to
change the thickness and color of every top-quark line separately: the
top-quark line is simply a separate object with well-defined properties.

The second consequence of the model-based approach is that \feyngame\ can
check whether a particular Feynman diagram is compatible with the interactions
of the underlying theory. Simply pressing the key \keystroke{f} is sufficient
to validate or disprove the Feynman diagram currently drawn on the canvas
corresponds as a physical amplitude.

The experienced particle physicist will scarcely need this feature, of
course. Indeed, its main application is educational in the context of the
``game mode'' of \feyngame. Currently, \feyngame\ offers one type of game,
named \texttt{InFin} (for \underline{In}itial and \underline{Fin}al), where
the initial and final state of an amplitude are generated (quasi-)randomly,
and the player has to construct a suitable valid (connected) Feynman diagram
within the underlying particle model. This game can be configured by the user
to a high degree, as will be described in more detail in \cref{sec:edu}.

%- }}}
%- {{{ section{General functionalities}

\section{General functionalities}

%- {{{ fig::frame

%
\begin{figure}
  \begin{center}
    \begin{tabular}{cc}
      \raisebox{0em}{%
        \mbox{%
          \includegraphics[%
            clip,width=.45\textwidth]%
                          {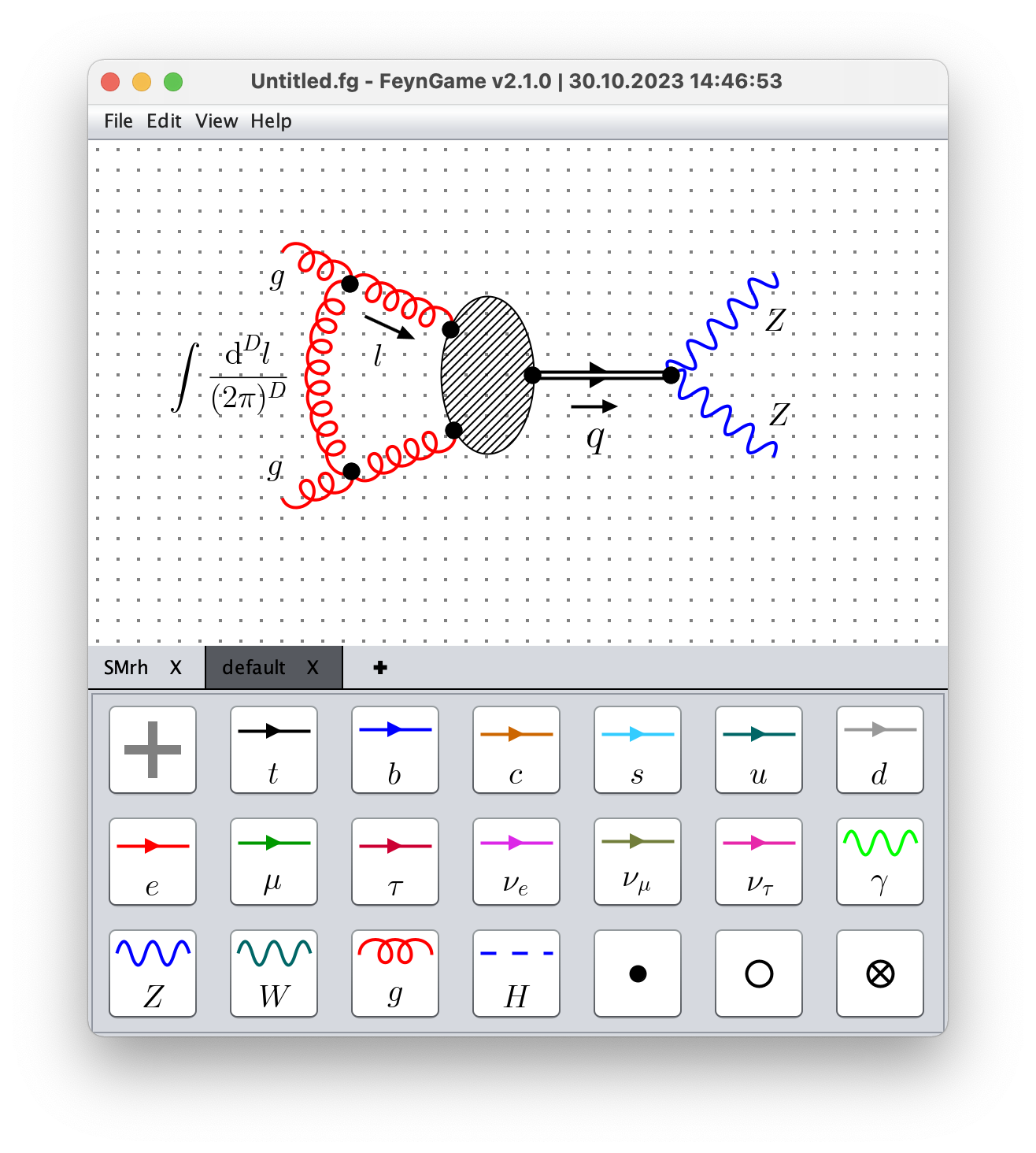}}}
      &
      \raisebox{0em}{%
        \mbox{%
          \includegraphics[%
            clip,width=.5\textwidth]%
                          {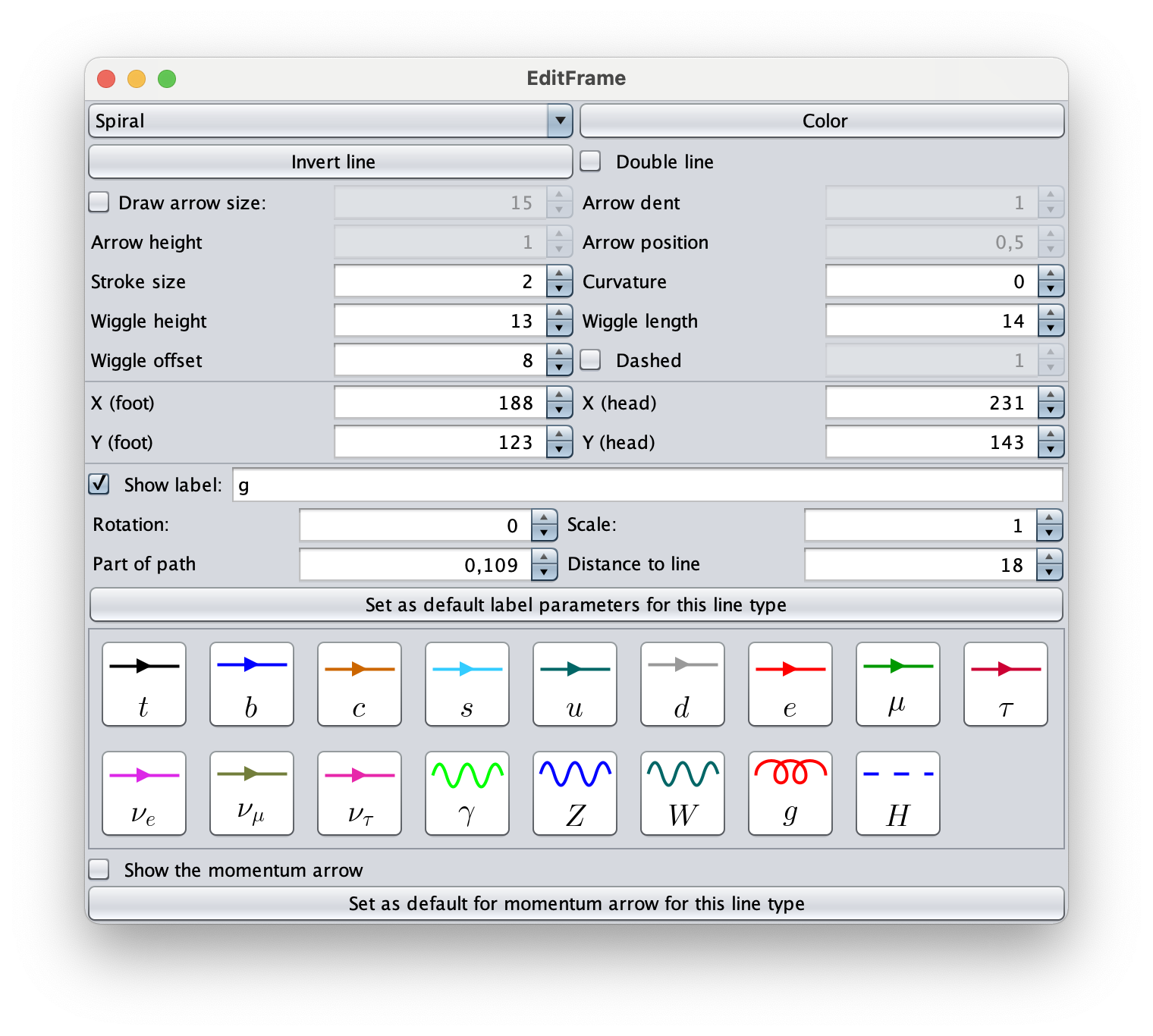}}}
    \end{tabular}
    \parbox{.9\textwidth}{
      \caption[]{\label{fig::frame}\sloppy Left: The \mainframe\ of
        \feyngame. The tiles in the lower part of the frame represent the
        \textit{current model}, the diagram itself is displayed on the
        canvas. The diagram shown here contains elements from the
        \sm\ (gluons, $Z$-bosons), but also non-standard elements like the
        hashed oval shape, and the double-line propagator which can be
        obtained via the menu items or the \editframe\ (right).}}
  \end{center}
\end{figure}
%

%- }}}

%- {{{ subsection{The main window}

\subsection{The main frame}

The left part of \cref{fig::frame} shows a screenshot of the so-called
\mainframe\ of \feyngame. The Feynman diagram in the upper part of
that window (the \textit{canvas}) has been drawn using a common input device
of the computer, such as the mouse or the trackpad, or most conveniently a
stylus device. For simplicity, we will refer to the input device as ``mouse''
in the following.

Below the canvas is a set of tiles, representing the lines (or particles) of
the \textit{current model}. Selecting one of these tiles allows one to
subsequently add the corresponding line to the canvas by clicking and
dragging. Trying this out, the user may soon discover a number of other quite
unique features of \feyngame:
\begin{itemize}
\item Initially, every line is drawn as a straight line. It can then be
  curved, for example by using the mouse wheel. The end points remain fixed
  during the curving operation.
\item If the end of one line is positioned close to a second line,
  \feyngame\ will connect the two through a vertex (no matter if it is
  compatible with the \textit{current model} or not). The second line will
  formally be split into two by this operation, such that the vertex actually
  connects three lines.
\item By clicking and dragging the lines on the canvas with the mouse, they
  can easily be moved (if clicked close to the line's center), rotated,
  stretched, or shortened (if clicked close to one of the line's ends).
\item Clicking and dragging a vertex will move the vertex and all the lines
  connected to it. This makes it very simple to modify the shape of a Feynman
  diagram.
\end{itemize}
The completed diagram can then be exported in many formats such as
\abbrev{JPG}, \abbrev{PDF}, or PostScript, which will be in one-to-one
correspondence to the image on the canvas (``what you see is what you get''),
modulo graphical aids like grid points and \textit{helper lines} which can be
activated in the representation on the canvas.\footnote{The export to
\abbrev{PDF} on MacOS sometimes does not correctly display the vertex
markers. We recommend to export to PostScript and subsequently convert to
\abbrev{PDF}, if desired.}  It can also be copied to the clipboard and thus
simply pasted to other applications such as Keynote or PowerPoint as
\abbrev{PNG} image (i.e.\ with transparent background) without the need of
saving the image to disk first.

%- }}}
%- {{{ subsection{The edit frame}

\subsection{The edit frame}

The \mainframe\ is sufficient for quick drawings where all lines have
their default attributes. Deviations from the default can be achieved for
example with the help of the \editframe, which opens by pressing
\keystroke{e}. The content of this window depends on the \textit{active}
object (an object can be \textit{activated} by clicking on it in the
canvas). For example, the right part of \cref{fig::frame} shows the
\editframe\ for one of the external gluon lines of the diagram shown on the
left. The \editframe\ allows one to modify all attributes of the active
object. For lines, this is the color, the width, the position on the canvas,
the curvature, the shape, the label, and much more. For vertex markers, one
can also choose a filling pattern, for example.

%- }}}

%- }}}
%- {{{ section{New drawing features}

\section{New drawing features}

Numerous features and improvements have been added to \feyngame\ since its
first release\,\cite{Harlander:2020cyh}. We highlight some of the most
important ones here. Some of them are exemplified in \cref{fig::frame}. A more
detailed list will be provided in a forthcoming publication.
\begin{description}[itemsep=0em]
\item[{\LaTeX\ labels}.]  \feyngame\ will interpret any text labels for lines
  or other graphical objects entered via the \editframe\ or in the \modelfile\
  as \LaTeX\ code.\footnote{\feyngame\ uses the Java library
  \href{https://github.com/opencollab/jlatexmath}{JLaTeXMath} for that
  purpose.} As usual, the canvas will display the compiled \LaTeX\ symbols in
  the same way as they would appear on the exported image.
\item[{Momentum arrows}.] For each line, \feyngame\ can draw an associated
  momentum arrow and optionally a momentum label. This feature is toggled by
  activating the line and pressing \keystroke{p}, or via the \editframe. The
  latter provides many options for the form and positioning of the momentum
  arrow and its label.
\item[Shapes.] In addition to lines and vertices, the graphical elements
  ``oval'' and ``rectangle'' are available. They can be inserted into the
  canvas via the menu item \menu[,]{Edit, Shapes}, and deformed and rotated as
  usual with the mouse. A full list of changeable attributes (fill color, fill
  pattern, line color, etc.) is obtained via the \editframe.
\item[Editing multiple objects.] By clicking a tile in the \textit{current
  model}, the attributes of that particular line type can be modified. This
  affects all current and subsequently drawn lines on the canvas in the
  current session. In order to adopt the changes also for future sessions, one
  can save the \modelfile\ using \menu[,]{File, Save model file}.
\item[Multiple model files.] One can work with several different models at a
  time and switch between them via tabs which appear between the canvas and
  the \textit{current model}.
\item[Model from diagram.] When loading a diagram that was drawn in an earlier
  session and then saved to disk, \feyngame\ can create a model which consists
  of all the lines and vertices which appear in that diagram. This can be
  helpful if the \modelfile\ which was used to produce the diagram is unknown,
  for example because the diagram was produced by someone else. Using the
  option to load several \modelfile s, it is then easy to add new lines to the
  diagram.
\item[Zoom and shift.] Diagrams can be moved as a whole on the canvas by
  pressing \keystroke{Shift} and click-dragging the canvas. One can also zoom
  into and out of the canvas with the arrow-up and arrow-down keys, or using
  the corresponding sub-items in \menu{View}.
\item[Performance improvements.] A number of issues related to \abbrev{CPU}
  efficiency or graphical representations have been improved or resolved.
\end{description}

%- }}}
%- {{{ section{New educational features}

\section{New educational features}\label{sec:edu}

So far, we have described new graphical features of \feyngame. However, also
the pedagogical part has been updated significantly.

%- {{{ subsection{Amplitude}

\subsection{Amplitude}

Provided that the \modelfile\ contains the required information on the Feynman
rules, \feyngame\ can produce the mathematical expression for the amplitude of
a particular (valid) Feynman diagram. Pressing \keystroke{f} will check the
diagram for validity, and if this is passed, it will open a dialog box where
the user can choose between seeing the formula for the amplitude, or copying
this formula in \LaTeX\ code to the clipboard which can simply be pasted into
any \LaTeX\ document.\footnote{Note that the expression for the amplitude can
be quite long and extend beyond the width of the screen. Also, \feyngame\ only
inserts the Feynman rules and does not attempt any simplifications on the
resulting expression.} The default model contains all the required information
for this. One can also ask \feyngame\ to display the momentum routing through
the diagram.

%- {{{ eeee

%
\begin{figure}
  \begin{center}
    \begin{tabular}{cc}
      \raisebox{0em}{%
        \mbox{%
          \includegraphics[%
            viewport=160 560 430 680,
            clip,width=.35\textwidth]%
                          {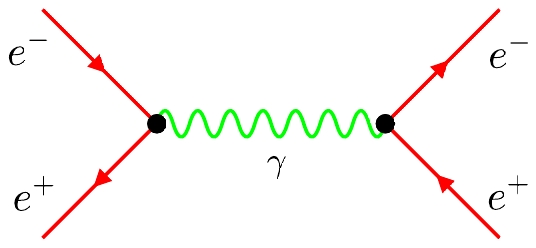}}}
      &
    \begin{minipage}[b]{.55\textwidth}
      {\footnotesize
      \begin{equation*}\label{eq::acyl}
  \begin{aligned}
&e^2\,u_{ \alpha_{1} }(\vec{p}_{1})\,\bar{v}_{ \beta_{1} }( (  - \vec{p}_{2} )
    )\,(-iQ_{e}\,\gamma_{ \beta_{1}  \alpha_{1} }^{ \mu_{1} })\,\\&
    \times i\left[\frac{-g_{
      \mu_{1}  \nu_{1} }}{ (  - p_{2} + p_{1} ) ^2+i\epsilon}
+
  (1-\xi_\gamma)\frac{ (  - p_{2} + p_{1} ) _{ \mu_{1} } (  - p_{2} + p_{1} )
  _{ \nu_{1} }}{( (  - p_{2} + p_{1} ) ^2)^2}\right]
    \,\\&\times
    (-iQ_{e}\,\gamma_{ \delta_{1}  \gamma_{1} }^{ \nu_{1} })\,v_{ \gamma_{1} }( (  - \vec{q}_{1} ) )\,\bar{u}_{ \delta_{1} }( (  - \vec{p}_{2} + \vec{q}_{1} + \vec{p}_{1} ) )\,    
  \end{aligned}
\end{equation*}}
    \end{minipage}
    \end{tabular}
    \parbox{.9\textwidth}{
      \caption[]{\label{fig::eeee}\sloppy When drawing the diagram on the left
        with \feyngame, it will produce the amplitude on the right.  }}
  \end{center}
\end{figure}
%

%- }}}

For example, if one uses the default model to draw the diagram for $e^+e^-\to
e^+e^-$ via the $s$-channel exchange of a photon shown in the left part of
\cref{fig::eeee}, and subsequently asks \feyngame\ for the amplitude, the
\LaTeX\ code will compile to the expression shown in the right part of
\cref{fig::eeee}.\footnote{The linebreaks were inserted manually.}

%- }}}
%- {{{ subsection{Level Generator}

\subsection{Level Generator}

The initial and final states generated in the \texttt{InFin} game mode are not
generated on the fly, but are read by \feyngame\ from a control file, the
so-called \textit{level file}. \feyngame\ comes with a default version of the
\textit{level file}, which is based on the \sm. It lists a number of possible
valid processes from which \feyngame\ randomly picks one and displays its
initial and final state particles on the canvas. The player is asked to
construct a connected Feynman diagram which mediates the process. If the task
is completed, Feyngame picks another pair of initial and final states from the
\textit{level file}.

If one wants to assume a different underlying particle model, one needs to
supply \feyngame\ with a corresponding \textit{level file}. Since constructing
such a file by hand can be quite tedious, in particular if the number of
processes to be played should be larger than just a few, \feyngame\ provides a
\textit{Level Generator}. Given a \modelfile, the maximal and minimal number
of particles in the initial and final state, and the maximal number of loops,
it uses the diagram generator
\texttt{qgraf}\,\cite{Nogueira:1991ex,Nogueira:2006pq} to generate the
requested number of processes and puts them into a \textit{level file}. We
refrain from a more detailed description of the level generator due to the
page limitations of these proceedings. It will be provided in a forthcoming
publication; the interested reader is welcome to contact us beforehand.

%- }}}

%- }}}
%- {{{ section{Plans for the future}

\section{Plans for the future}

As pointed out above, the features described in this paper are a collection of
just some of the most important changes since the original release of
\feyngame, and the interested reader is invited to try out the new version. In
addition, many other features are planned, in preparation, or even already
contained in preliminary versions. Among the latter is the automatic
visualization of \texttt{qgraf} output, which should be a very useful feature
for debugging code that calculates Feynman diagrams. There are also more
sophisticated versions of \texttt{InFin}, where, in addition to the given
initial and final state particles, also specific internal lines have to be
incorporated in the final Feynman diagram. Future releases of
\feyngame\ should also allow for editing groups of lines in a Feynman diagram
simultaneously.

%- }}}
%- {{{ section{Conclusions}

\section{Conclusions}

We have presented a number of new features which are available in the latest
version of \feyngame\ and hope that they will be helpful to the physics
community in producing high-quality publication-level Feynman diagrams in an
efficient way. We also hope that \feyngame\ helps to convey the structure of
Feynman diagrams to particle physics novices.

Any feedback or requests for help or new features are welcome and can be sent
directly to the authors, or submitted via the
\href{https://gitlab.com/feyngame/FeynGame}{\texttt{gitlab}} repository. 

%- }}}
%- {{{ Acknowledgments}

\paragraph{Acknowledgments.}

We would like to thank all users of \feyngame\ for their feedback. Special
thanks go to Jakob Linder for pointing out a number of issues, Lars B\"undgen
and Erik de la Haye for contributing to the \feyngame\ development, and
Maximilian Lipp for designing \feyngame's structure in such a flexible way.

%- }}}

\providecommand{\href}[2]{#2}\begingroup\raggedright
\endgroup


\begin{thebibliography}{1}

\bibitem{Harlander:2020cyh}
R.~V. Harlander, S.~Y. Klein, and M.~Lipp, {\emph{FeynGame}},
  \href{http://dx.doi.org/10.1016/j.cpc.2020.107465}{{\em Comput. Phys.
  Commun.} {\bfseries 256} (2020) 107465},
  \href{http://arxiv.org/abs/2003.00896}{{\ttfamily arXiv:2003.00896
  [physics.ed-ph]}}.

\bibitem{Nogueira:1991ex}
P.~Nogueira, {\emph{Automatic Feynman graph generation}},
  \href{http://dx.doi.org/10.1006/jcph.1993.1074}{{\em J. Comput. Phys.}
  {\bfseries 105} (1993) 279--289}.

\bibitem{Nogueira:2006pq}
P.~Nogueira, {\emph{Abusing qgraf}},
  \href{http://dx.doi.org/10.1016/j.nima.2005.11.151}{{\em Nucl. Instrum. Meth.
  A} {\bfseries 559} (2006) 220--223}.

\end{thebibliography}
\end{document}